\documentclass[sigconf,authorversion,nonacm]{acmart}

\AtBeginDocument{%
  \providecommand\BibTeX{{%
    \normalfont B\kern-0.5em{\scshape i\kern-0.25em b}\kern-0.8em\TeX}}}

\setcopyright{acmcopyright}
\copyrightyear{2021}
\acmYear{2021}
\acmDOI{10.1145/1122445.1122456}

\acmConference[SPLC’21]{25th ACM International Systems and Software Product Lines Conference}{06–11 September, 2021}{Leicester, UK}
\begin{document}

\title{Applying multi product lines to equity market software ecosystem}

\author{Khosro Pakmanesh}
\affiliation{%
  \institution{Ferdowsi University of Mashhad}
  \streetaddress{Azadi Square}
  \city{Mashhad, Razavi Khorasan Province}
  \country{Iran}}
\email{khosro.pakmanesh@mail.um.ac.ir}

\author{Mehdi Mojaradi}
\affiliation{%
  \institution{University of Tehran}
  \streetaddress{Enghelab Square}
  \city{Tehran, Tehran Province}
  \country{Iran}}
\email{mehdi.mojaradi@ut.ac.ir}

\begin{abstract}
\textit{Context:} 
In recent decades, many financial markets and their participants have changed their working method from a completely manual and traditional one to an automatic one, benefiting from complex software systems. There are different approaches to the development of such software systems. \\
\textit{Objective:} 
In this paper, we study the application of the Multi Product Line (MPL) approach in the software ecosystem (SECO) of the equity market. By using the concepts and practices of the MPL approach, we want to design a SECO that makes the automated flow of financial transaction data between market participants' software pieces possible. \\
\textit{Method:}
We use the software product line engineering paradigm for the research method. This paradigm divides the engineering process into domain and application engineering processes. Finally, we discuss three order life cycle scenarios by which the derived products are tested and validated.\\
\textit{Results:}
To implement the mentioned working method, named Straight-through Processing (STP), different technical and non-technical elements' contribution is essential. Attaining success in developing the equity market SECO addresses the technical aspect and prepares the technical infrastructure for the rest of the work. \\
\textit{Conclusion:}
The successful validation of the equity market SECO indicates that the adoption of the MPL approach is a viable strategy for the development of equity market SECOs. It also suggests that this approach is worthy of more attention and investment.
\end{abstract}




\maketitle

\section{Introduction}
In the past, the trade settlement process with traditional means usually used to take three days in the United States. Over time, with more pressure from US financial authorities, the trade settlement's required time was reduced to one day. However, this was not the end of the line. This improvement gradually converted to a paradigm for automating all activities among market participants to settle trades \cite[p.~17]{Khanna2008straight}. The STP is the name given to business processes and the technical infrastructure making this paradigm possible \cite[p.~1]{Khanna2008straight}. Khanna mentions seven market trends that forced the traditional financial markets to move towards the STP \cite[pp.~2-17]{Khanna2008straight}. 

The most significant pillars of STP are easy communication between market participants, real or near real-time information processing, automated instead of manual processing of information, and concurrent information exchange \cite[pp.~17-18]{Khanna2008straight}. STP has also provided its participants with enormous advantages. Some of the most notable ones are shortened settlement time, minimum human intervention, standardized communication protocols, and increased throughput leading to higher transaction numbers \cite[pp.~29,30]{shetty2006practical}.

Bosch classifies the technical infrastructure or, in other words, the software system paving the way for achieving STP as a SECO for its particular properties \cite{bosch2009software}. There are some considerable technical and non-technical challenges associated with SECOs \cite{jansen2009sense} \cite{joshua2013software}. Responding to these new challenges with ad-hoc solutions might cause software development companies to lose a fortune.

An interesting property of equity market SECOs is their similar foundations despite their implementation differences. Market participants, such as custodians, brokers, exchanges, clearing banks, the clearing corporation, are usually seen across different equity markets with reasonably similar functionalities and dedicated software. 

Clements and Northrop describe Software Product Lines (SPLs) as an approach bringing the \textit{economies of scope} to their users. To put it in simple words, an SPL owner can reap the economic benefit of the similarities of several products \cite[p.~9]{clements2002software}. As a result, the mentioned property makes the SPL approach a suitable candidate for guiding the development of each market participant's software. Then, these market participants or members, as a SECO, can take advantage of the MPL approach to manage their relationships.

An \textit{SPL} is simply a product line that consists of some software products sharing a set of features to cater to a particular domain's needs \cite[p.~5]{clements2002software}. In this situation, each software product is derived from a set of shared features in a predefined manner. An \textit{MPL} is a set of independent software product lines that together conduct a specific activity and form a large-scale system \cite{holl2012systematic}.

In this paper, we intend to apply the MPL approach to the equity market SECO. At the first stage, in the domain engineering section, we provide background information about the equity market, its participants, and their relations. We also analyze the variability in each market participant's software. Next, we describe the employed architecture. In the application engineering section, we discuss the implementation and derivation of two SECOs. Finally, we test the derived SECOs to validate them and also demonstrate that our MPL has the necessary foundations for establishing STP.

\section{Related Work}
As the ultimate goal of this paper is to study the application of the MPL approach in a subdomain of the financial domain, we investigated the SPL approach's application in other subdomains of the financial domain to learn more about other work. Next, since increasing Software Reuse (SR) is one of the most important foundations of the SPL approach, and this proposed case study is recognized as a SECO, we inspected the SR concept in the context of SECOs. Finally, to show an effective and successful application of the MPL approach in managing and organizing a SECO, we reviewed an advertising SECO. 

Regarding the application of the SPL approach in the financial domain, there are different papers that can be associated with different phases of the Software Development Life Cycle (SDLC). We start our review with Ihme's work. Ihme studied the application of the SPL and agile application development approaches in large and complex financial IT systems by interviewing different roles in a financial company's software development department \cite{ihme2013scrum}.

Krsek et al. reported some observed obstacles during the SPL adoption in two large financial service organizations whose main business was not software. Then, they introduced some mechanisms to make the SPL adoption easier \cite{krsek2008experiences}. Witman examined the adoption of the SPL approach and its practices in a large financial institution. The author also discussed two different SPL maturity levels and their conflicting goals with the time-to-market requirement for projects \cite{witman2009software}. Verlage and Kiesgen described that the Market Maker Software AG company used the SPL approach to create a financial SPL. To reach this objective, some changes in business and the organization were made. This paper reiterated the learned lessons of the five-year software adoption journey of this company \cite{verlage2005five}. Quilty and Cinnéide scrutinized the gradual conversion of a single system into an SPL. In this paper, the authors stated that the owner company had introduced three principles to prevent regular SPL adoption challenges. Finally, according to gathered data over a ten-year period, they illustrated some achievements attributed to SPL adoption \cite{quilty2011experiences}. 

Altintas et al., by profiting from the SPL's concepts and practices, introduced a domain specific kit abstraction for design and implementation of a family of financial gateway systems and then presented the result of their experimentation \cite{altintas2008octopoda}. Jeong Ah Kim and SunTae Kim employed the SPL's architecture in creating an e-insuBanking product line. With this architecture and also a customizing environment, they succeeded in deriving two different products \cite{kim2014case}. 

Altintas et al. introduced a financial SPL named Aurora and used it as a platform to derive different financial products. They discussed the SPL's fundamental activities in the context of the Aurora. They also discussed other topics, such as development environments, life cycle management practices, and quality assurance tools \cite{altintas2005aurora}. Ye et al. studied Wingsoft financial management systems SPL and investigated the impact of features on the SPL's components. They also scrutinized the virtues and flaws of different variation mechanisms and defined some criteria for their selection \cite{ye2009case}.

Jørgensen et al. described the development of mobile banking apps with regards to variability management and improving SR. They also reported their challenges and learned lessons during the development of software for iOS and Android platforms \cite{jorgensen2016variability}.

Concerning the concept of SR in the context of SECOs, Santos presented a method to define, model, and analyze SECOs by using reuse-driven techniques and other non-technical concepts \cite{santos2014reuseseem}.

Urli et al. show the MPL approach's efficacy in organizing a medium-scale advertising SECO by relying on a case study named the YourCast project \cite{urli2014managing}. They believe that adopting the MPL approach in their SECO has led to the separation of concerns, facilitating the management activities for an ever-expanding community, preserving the consistent interaction of artifacts, and supplying the end-users with features to create their desired products.

According to the reviewed work, the MPL approach has not previously been used for the development of the equity market SECO to the best of our knowledge.

\section{Research Method}
This paper uses Pohl et al.'s Software Product Line Engineering (SPLE) paradigm\cite{pohl2005software} for its research method. This paradigm separates the \textit{domain engineering} and \textit{application engineering} processes. The Domain engineering process is concerned with defining an SPL's commonality and variability. The application engineering process in SPLE is the process in which the product(s) are derived from an SPL's core assets by benefiting from the SPL's variability. 

Both domain and application engineering processes contain sub-processes to separate concerns. For brevity, we discuss \textit{domain requirements engineering} and \textit{domain design} sub-processes from the domain engineering process and \textit{application realization} sub-process from the application engineering process.

\section{Domain Requirements Engineering}
In this section, we will first describe how we extracted the equity market SECO's domain knowledge from some sources. After that, we explain how we modeled the equity market SECO's domain knowledge. Then, we introduce the participants of an equity market SECO and their relations by using A diagram. Following that, we study variability in the equity market SECO by drawing each market participant's feature diagram and illustrating a feature diagram's variation points.

\subsection{Domain knowledge extraction}
We used different sources to extract the equity market's domain knowledge and the most important transactions in this type of market. Before introducing the sources, an important point should be illustrated. Since there were different interpretations of the activities happening in the equity market and its transactions, each of which had slight differences, we had to choose one of them for the entire paper to avoid unwanted ambiguities.

To begin introducing the sources we have used, we should start with our primary source. Shetty and Jayaswal described the fundamental concepts of the equity market, including the equity market's activities and two of its most important transactions, from initiating the order to its settlement \cite[pp.~1-30, 41-47, 299-306]{shetty2006practical}. As the secondary source, Khanna did the same with this difference that she explained the same concepts and transactions in the context of the United States's financial markets \cite[pp.~21-50]{Khanna2008straight}. 

Since we wanted to study the domain more generally and avoid any bias in favor of any implementations done in different countries, we decided to choose the former source as the main source, describing the concepts without relying on a particular implementation. However, the secondary source shed light on some vague aspects of the transactions' processes and contributed remarkably to gaining a deeper understanding.

\subsection{Domain Modeling}
To model the domain, we read the textual information from sources and extracted entities and processes. We then used this knowledge to form a simplified object-oriented model of the system, where we can apply the MPL approach and benefit from all virtues it brings along. Because creating a real-world SECO demands many experts in different fields and encompasses a wide range of topics, we intentionally decided to limit the software's scope and concentrate on parts of this SECO that are important to us. To do so, we left out all details related to the fees that a participant should pay to another. Moreover, since there are different algorithms in each market participant's software that requires various experts, we decided not to implement many complicated algorithms and replace them with fake ones, acting as a placeholder to illustrate where the real algorithms are used.

As you can see in the next sections, different market participants, such as custodians, brokers, exchanges, clearing banks, the clearing corporation, and the depository, have different software in the equity market SECO. Although this paper covers all of these participants and even brings them in its SECO's source code\footnote{\href{https://github.com/khosropakmanesh/equitymarketsoftwareecosystem} https://github.com/khosropakmanesh/equitymarketsoftwareecosystem}, it mainly concentrates on the most important ones, namely custodians, brokers, exchanges, and the clearing corporation. Needless to say, clearing banks and depositories are not considered as key parts of an equity market SECO because of the infrastructural role they play in transactions. Usually, they are provided for the SECO as external services.

\subsection{Equity market's participants and their relations}
Before discussing each market participant's software and its variability in detail, it is essential that we provide a brief definition of each participant and then depict the relation of participants with each other. 

A \textit{broker} is an organization that connects clients with exchanges by routing the client's orders to exchanges. It also provides them with some services to help them have a successful trade. A broker mainly has two types of clients, \textit{retail and institutional} ones. For institutional clients, such as fund managers, another organization called \textit{custodian} acts as an agent for safekeeping their money or equities and protecting their interests. Custodians help institutional clients with the activities of clearing and settlement. \textit{Exchange} is a place where orders are matched, and trades are made. A \textit{clearing corporation} is responsible for the activities of clearing and settlement of trades. It guarantees that the two sides of a trade fulfill their commitments. \textit{Clearing banks}, which are the members of clearing corporations, keep the money of the buy-side of a trade. A \textit{Depository} does the same for the sell-side of a trade by keeping its equities. 

\begin{figure}[h]
  \centering
  \includegraphics[width=\columnwidth]{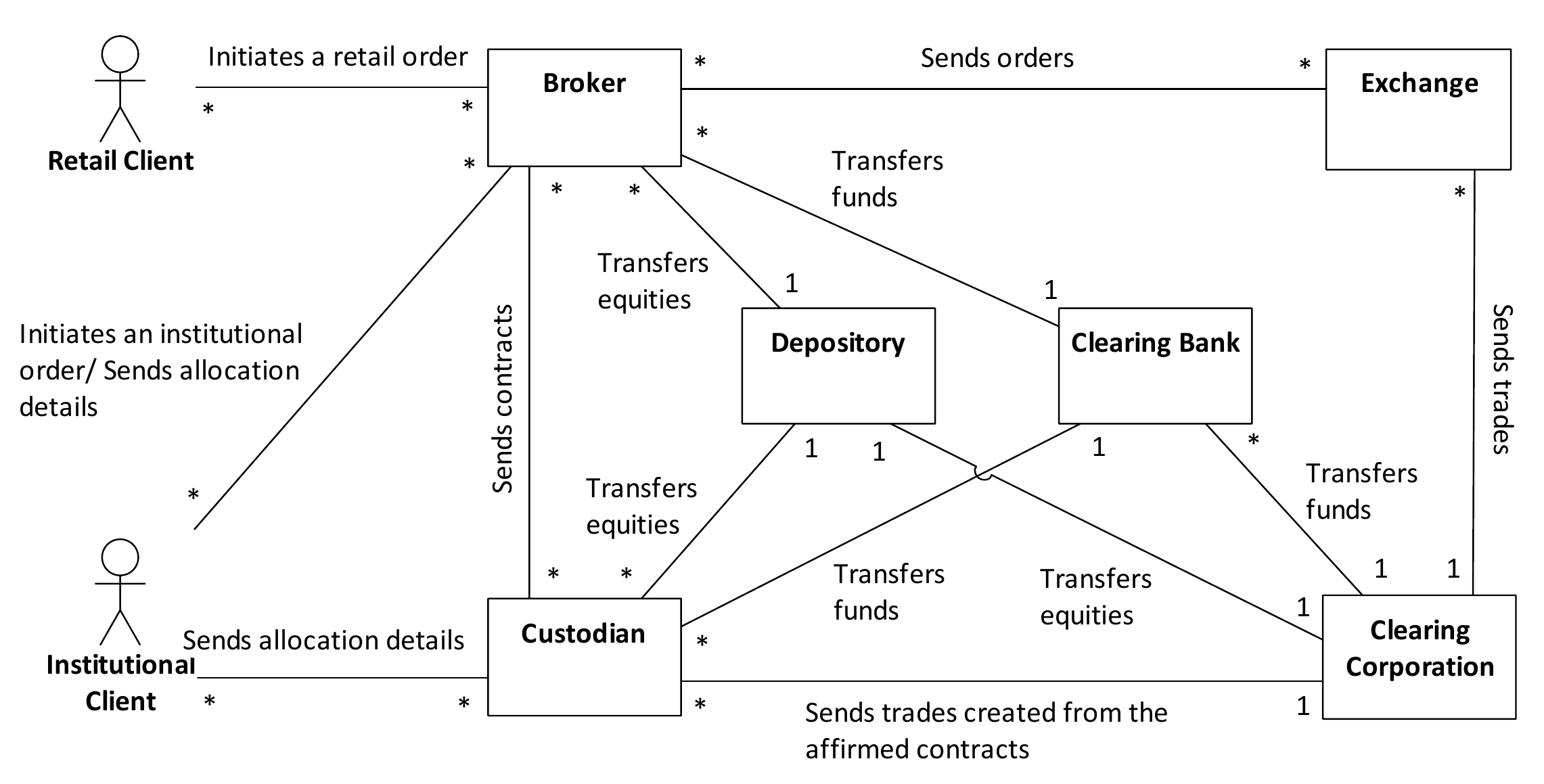}
  \caption{Equity market's participants and their relations}
  \Description{Note that both clients interact with the system only through brokers and custodians.}
  \label{fig:participantsRelations}
\end{figure}

Now that we generally know the responsibility of each unit, we can better understand their relations. The relation of different participants has been delineated in figure~\ref{fig:participantsRelations}.

\subsection{Variability in equity market}
In this section, we show the variability in each market participant's software. We use the FeatureIDE software plugin for modeling of features \cite{leich2005tool}. FeatureIDE is an IDE based on the Eclipse IDE tailored to meet the development requirements of SPLs. This software plugin covers all stages of feature-oriented software development \cite{featureide}. At this stage, we use the modeling feature of the FeatureIDE software plugin to model the features. 

Before going further, it is necessary to review two concepts of \textit{variation points} and \textit{variants}. A variation point can be considered as a placeholder where different variations of a specific functionality can be placed. Variants are actually the variations of the mentioned functionality \cite[p.~115]{clements2002software}. 

Note that in FeatureIDE, abstract features, colored in light purple, have been used for two purposes: firstly, to indicate the classifications of variants and, secondly, to specify the variation points. Concrete features, which have been colored in violet, represent the variants.

In the development process of the equity market SECO, 36 variation points and 100 variants have been identified in total. Since describing each variant is out of the scope of this paper, we decided to just describe the variation points and refrain from describing the variants.

\subsection{Market participants' SPLs and their variation points}
\begin{figure}[h]
  \centering
  \includegraphics[width=\columnwidth]{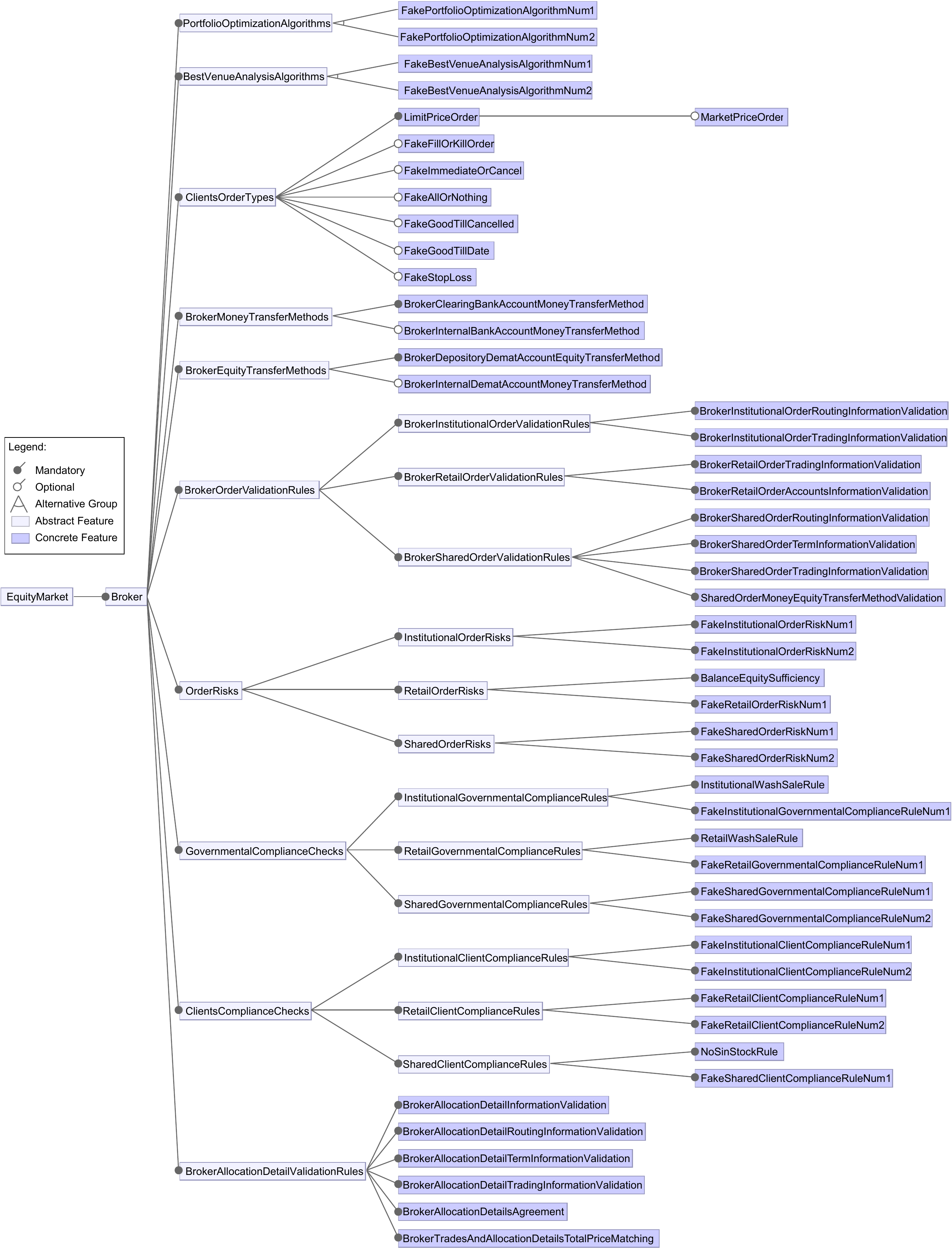}
  \caption{Broker feature diagram}
  \label{fig:brokerFeatureDiagram}
\end{figure}

\begin{figure}[h]
  \centering
  \includegraphics[width=\columnwidth]{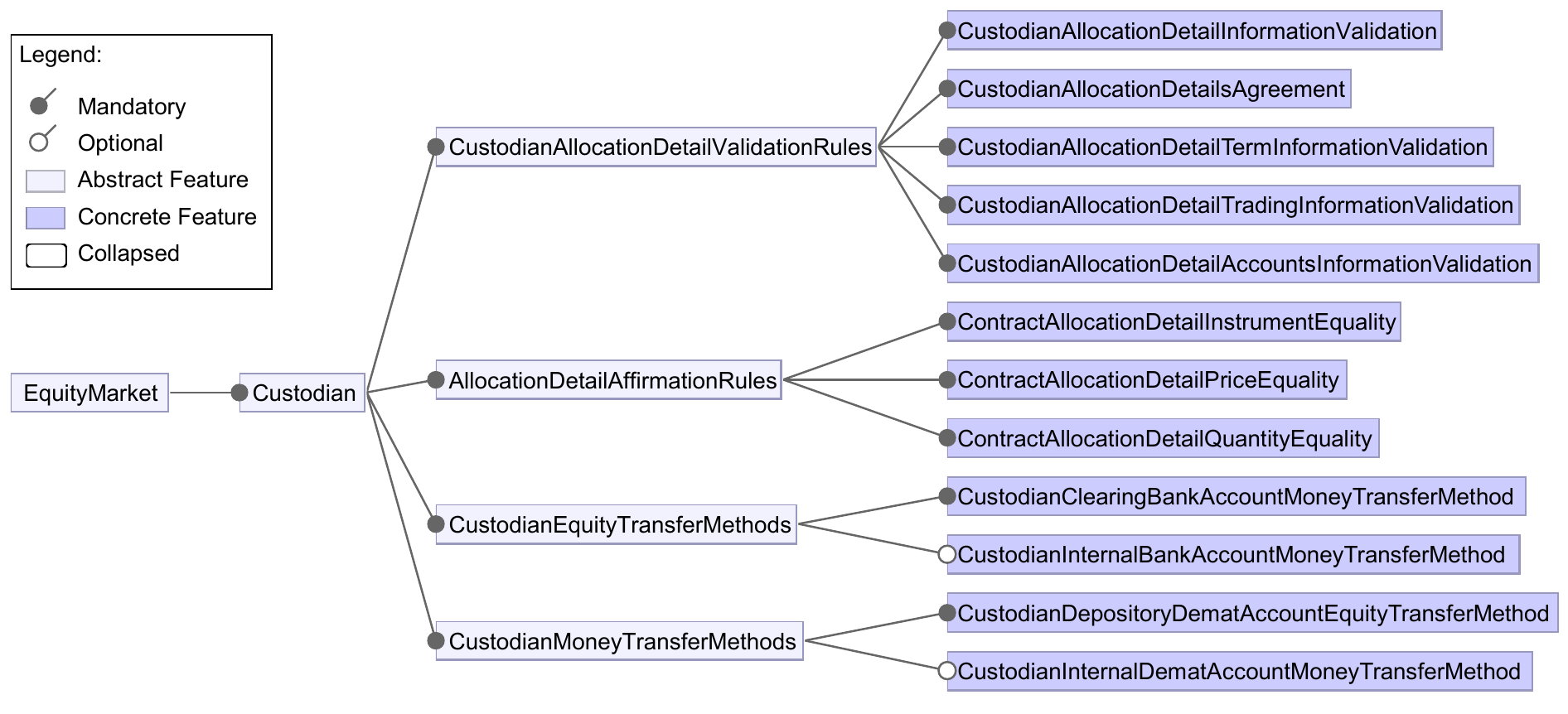}
  \caption{Custodian feature diagram}
  \label{fig:custodianFeatureDiagram}
\end{figure}

\begin{figure}[h]
  \centering
  \includegraphics[width=\columnwidth]{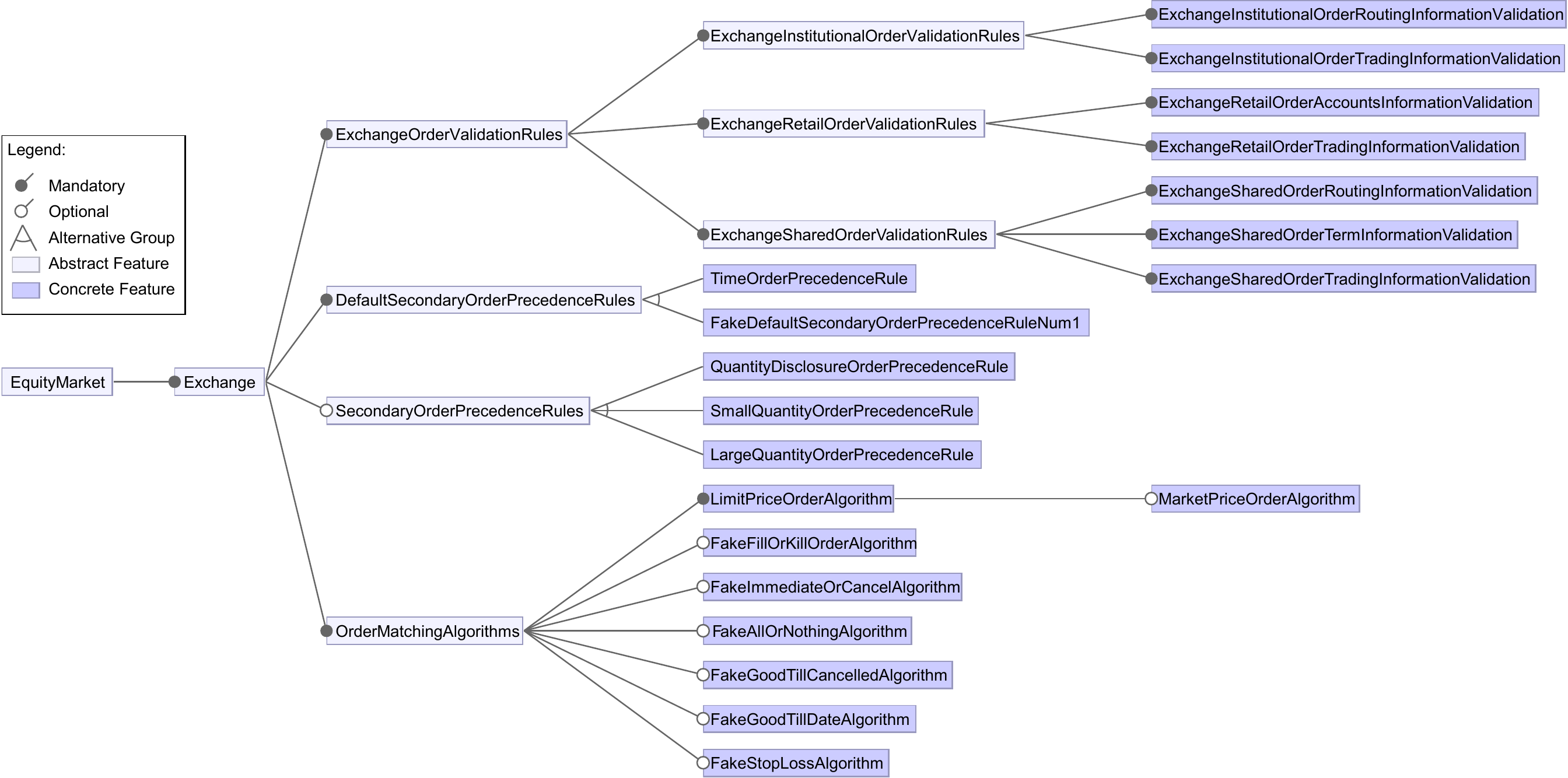}
  \caption{Exchange feature diagram}
  \label{fig:exchangeFeatureDiagram}
\end{figure}

\begin{figure}[h]
  \centering
  \includegraphics[width=\columnwidth]{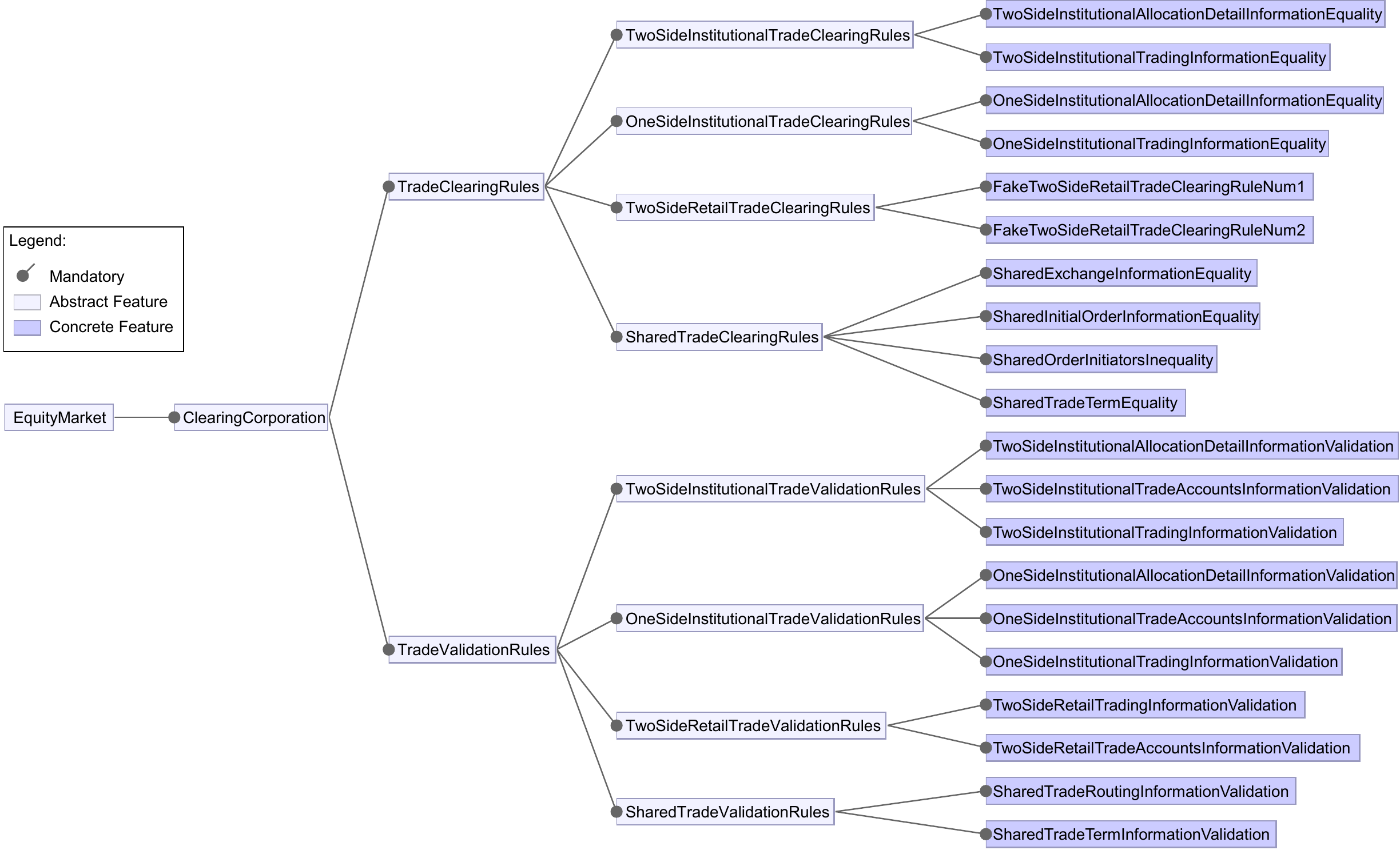}
  \caption{Clearing corporation feature diagram}
  \label{fig:clearingCorporationFeatureDiagram}
\end{figure}

The feature diagrams of the broker SPL, custodian SPL, exchange SPL, and the clearing corporation SPL have been shown in figures~\ref{fig:brokerFeatureDiagram}, ~\ref{fig:custodianFeatureDiagram}, ~\ref{fig:exchangeFeatureDiagram}, and ~\ref{fig:clearingCorporationFeatureDiagram} respectively. The variation points of each software piece have been detailed in tables ~\ref{tab:brokerVariationPoints}, ~\ref{tab:custodianVariationPoints}, ~\ref{tab:exchangeVariationPoints},  and ~\ref{tab:clearingCorporationVariationPoints} respectively. Note that we formed each market participant's feature model in the domain engineering stage by studying similarities and differences among varied domain concepts.

\begin{table*}
  \centering
  \caption{broker software's variation points}
  \label{tab:brokerVariationPoints}
  \begin{tabular}{|p{3.5cm}|p{13.5cm}|}
    \hline
    Variation point name&Description\\
    \hline
    Order validation rules&When either institutional or retail clients initiate their orders, at the first stage, some basic order validation rules should be checked to ensure the validity of input data.\\
    \hline
    Portfolio optimization algorithms&Portfolio optimization is the activity of choosing the best portfolio from a set of possible portfolios. Either retail or institutional clients can delegate this activity to a broker. This task can be done by different algorithms, each with a specific strategy.\\
   \hline
    Best venue analysis algorithms&Either retail or institutional clients can delegate the responsibility of choosing an execution venue (exchange) to a broker. Some criteria, such as the best price, the execution speed, and the lowest price of transactions, play an essential role in decision making. This task can be done by different algorithms, each with a specific strategy.\\
   \hline
       Client's order types&When either institutional or retail clients initiate their orders, they might want to define some rules to guide the exchange's matching engine for the order matching process. Shetty and Jayaswal describe some of these order types in their book \cite[pp.~43,44]{shetty2006practical}\\
   \hline
       Broker money transfer methods&When a buy-side retail client initiates an order to buy some equities, they need a payment method to transfer the money required for buying equities to the broker. Similarly, when a sell-side retail client initiates an order to sell some equities, they need a money receive method to receive the money from the broker after the settlement phase. These activities can be done with different money transfer methods.\\
   \hline
       Broker equity transfer methods&As we mentioned for the buy case, both buy and sell-side retail clients need a method to transfer equities. The activities of transferring equities for both sides can be done with different equity transfer methods.\\
   \hline
	       Order risks&When either institutional or retail clients initiate their orders, some risks or issues should be managed before sending an order to the exchange.\\
   \hline
       Governmental compliance checks&Government issues some rules to restrict illegal actions in financial markets. When either a retail or institutional client initiates an order, these rules should be checked.\\
   \hline
       Client's compliance checks&Sometimes, either a retail or institutional client delegate the responsibility of defining some aspects of an order to a broker. In such cases, they provide the broker with some general instructions to determine these aspects.\\
   \hline
          Broker allocation detail validation rules&Once an institutional client enters a trade's allocation details, basic allocation detail validation rules should be checked to ensure the validity of input data.\\
   \hline
\end{tabular}
\end{table*}

\begin{table*}
  \centering
  \caption{Custodian software's variation points}
  \label{tab:custodianVariationPoints}
  \begin{tabular}{|p{3.5cm}|p{13.5cm}|}
    \hline
    Variation point name&Description\\
    \hline
    Custodian allocation detail validation rules&Once an institutional client benefiting from a custodian's services enters a trade's allocation details, basic allocation detail validation rules should be checked to ensure the validity of input data.\\
    \hline  
    Allocation detail affirmation rules&When a custodian receives contracts from a broker and makes sure they are valid, it is time to analyze them against their respective allocation details. There are some rules that a custodian uses to ensure that the contracts and allocation details are in agreement.\\
    \hline   
     Custodian money transfer methods&When a buy-side institutional client deposits its money with a custodian to pay for the bought equities in the settlement stage, they need a payment method to do this activity. Similarly, when a sell-side institutional client wants to receive the money gained from the trade, they need a money receive method to receive the money from its custodian after the settlement phase. These activities can be done with different money transfer methods.\\
    \hline   
    Custodian equity transfer methods&Similar to what we mentioned for the buy case, both buy and sell-side institutional clients need a method to transfer equities. The activities of transferring equities for both sides can be done with different equity transfer methods.\\
    \hline   
\end{tabular}
\end{table*}

\begin{table*}
  \centering
  \caption{Exchange software's variation points}
  \label{tab:exchangeVariationPoints}
  \begin{tabular}{|p{3.5cm}|p{13.5cm}|}
    \hline
    Variation point name&Description\\
    \hline
    Exchange order validation rulese&When a broker sends an order to an exchange, at the first stage, some basic order validation rules should be checked to ensure the validity of input data. \\
    \hline  
    Secondary order precedence rules&The order precedence rules are the rules that an exchange's order matching engine uses to rank orders for trading. The order that has a higher rank takes more priority for matching. The price of an order is the primary precedence rule. After that, markets have different secondary precedence rules.\\
    \hline  
    Default secondary order precedence rules&If there are some equal orders based on the market's secondary order precedence rule, the default secondary order precedence rule determines their trading order.\\
    \hline  
    Order Matching Algorithms&The exchange's order matching engine uses the order type that either an institutional or retail clients have defined for the broker. An exchange has a peer algorithm for each order type in its order matching engine. Therefore, at the order initiation phase, a client has to choose from order types supported by the exchange.\\
    \hline  
\end{tabular}
\end{table*}

\begin{table*}
  \centering
  \caption{Clearing corporation software's variation points}
  \label{tab:clearingCorporationVariationPoints}
  \begin{tabular}{|p{3.5cm}|p{13.5cm}|}
    \hline
    Variation point name&Description\\
    \hline
    Trade validation rules&When an exchange or a custodian send a trade to a clearing corporation, at the first stage, some basic trade validation rules should be checked to ensure the validity of input data.\\
    \hline  
    Trade clearing rules&A Clearing corporation uses different clearing rules in the clearing stage of trades.\\
    \hline  
\end{tabular}
\end{table*}

\subsection{Cross-tree constraints}
\begin{figure}[h]
  \centering
  \includegraphics[scale=0.75]{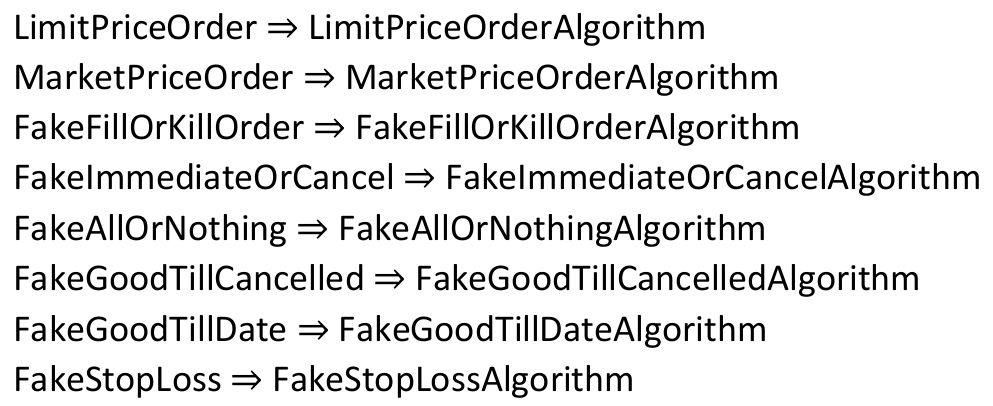}
  \caption{Equity market's cross-tree constraints}
  \label{fig:cross-treeConstraints}
\end{figure}

In featureIDE when we cannot define the constraints between features in a hierarchical structure, we use cross-tree constraints to do so. An example of this situation is when we want to show that a feature from a part of a feature diagram depends on another feature from another part of the same diagram. FeatureIDE uses propositional formulas to depict these relations. In our case, we use cross-tree constraints to indicate that the existence of each order type depends on the existence of its respective order matching algorithm. The cross-tree constraints of the equity market's feature diagram have been delineated in figure ~\ref{fig:cross-treeConstraints}.

\section{Domain Design}
In this section, we describe the equity market SECO's architecture and its limitations.

\subsection{Equity market's architecture}
\begin{figure}[h]
  \centering
  \includegraphics[width=\columnwidth]{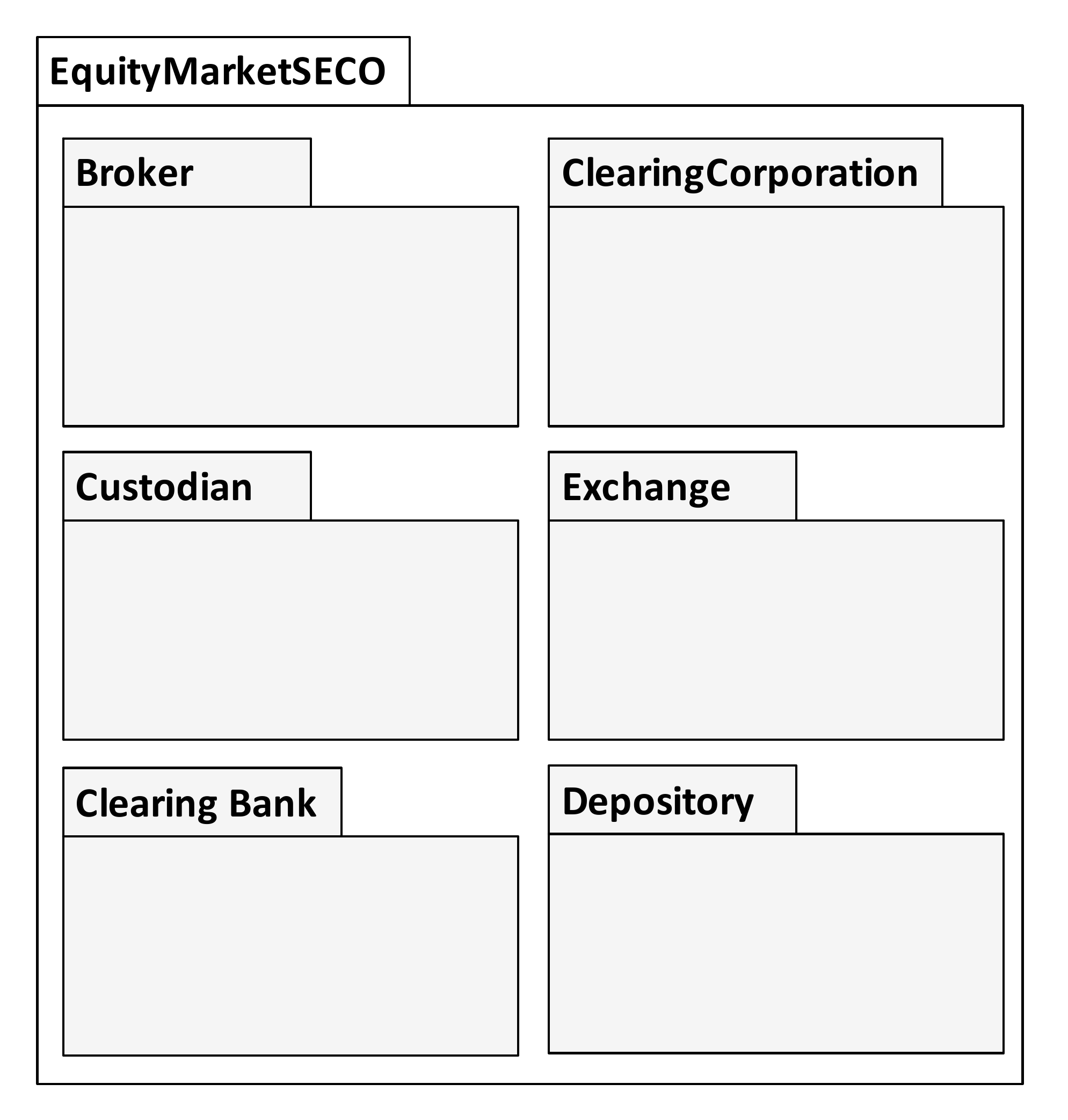}
  \caption{Equity market's software architecture}
  \Description{}
  \label{fig:emSECOArc}
\end{figure}

\begin{figure}[h]
  \centering
  \includegraphics[width=\columnwidth]{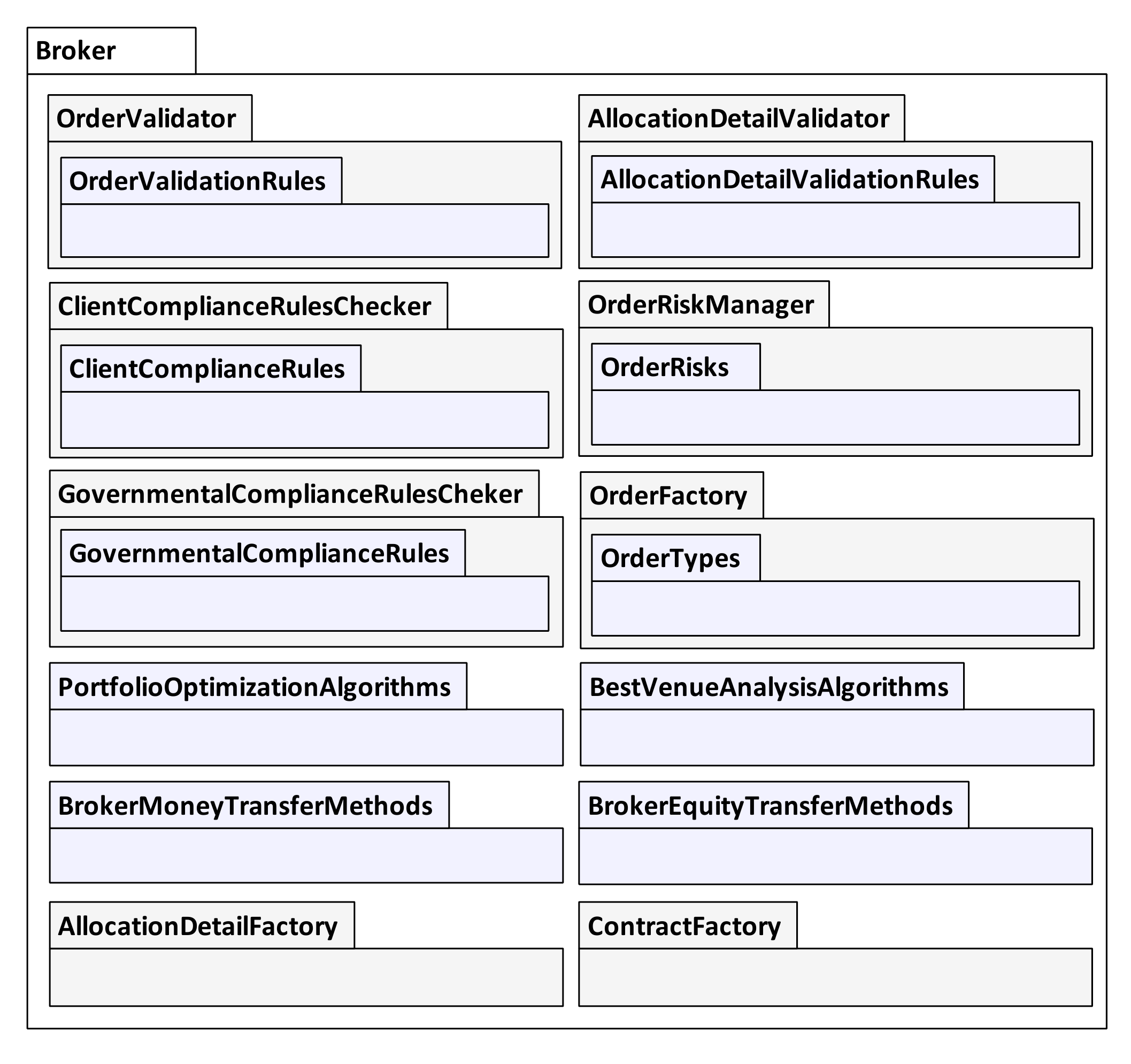}
  \caption{Broker's software architecture}
  \Description{}
  \label{fig:brokerArc}
\end{figure}

\begin{figure}[h]
  \centering
  \includegraphics[width=\columnwidth]{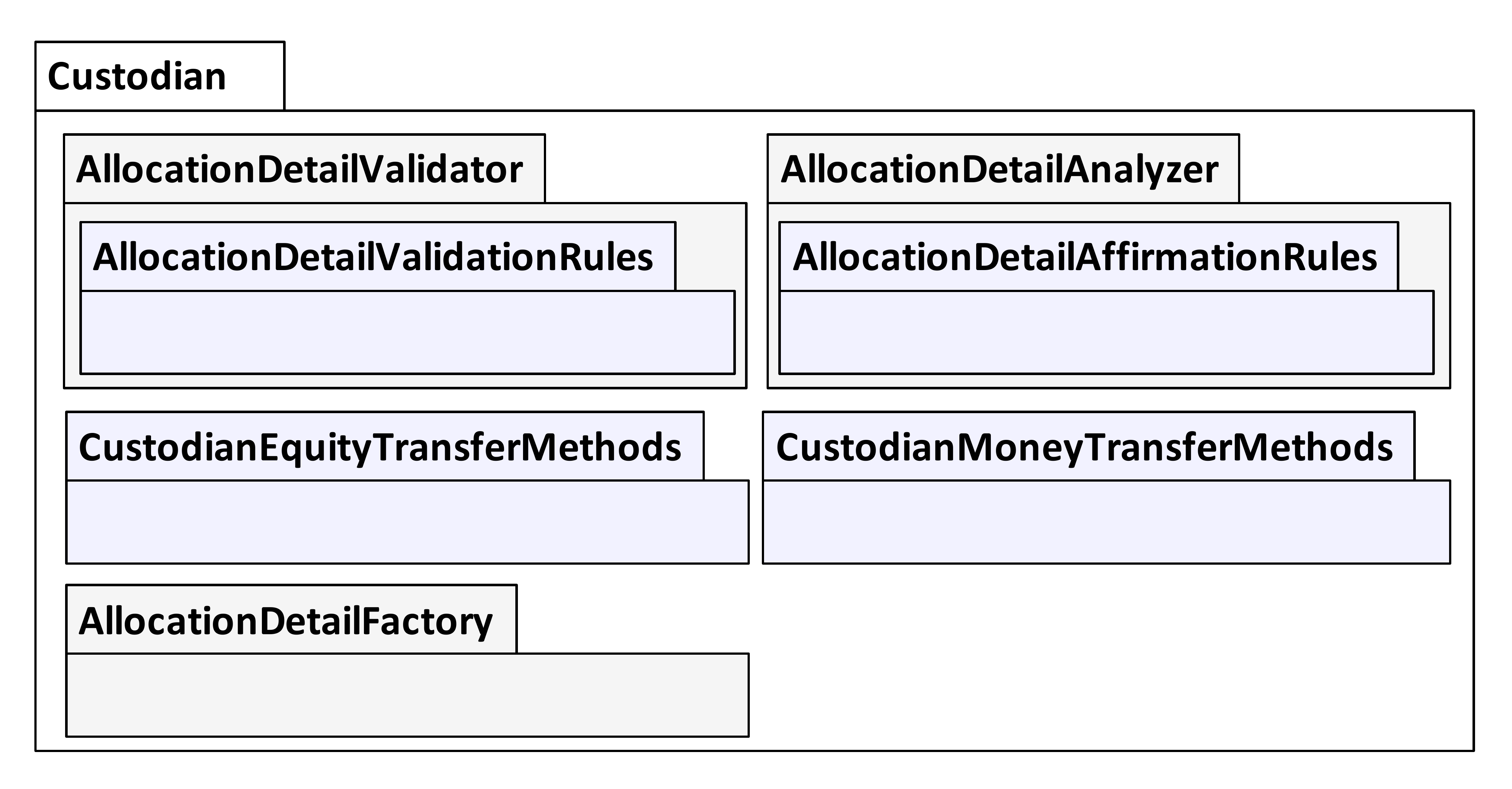}
  \caption{Custodian's software architecture}
  \Description{}
  \label{fig:custodianArc}
\end{figure}

\begin{figure}[h]
  \centering
  \includegraphics[width=\columnwidth]{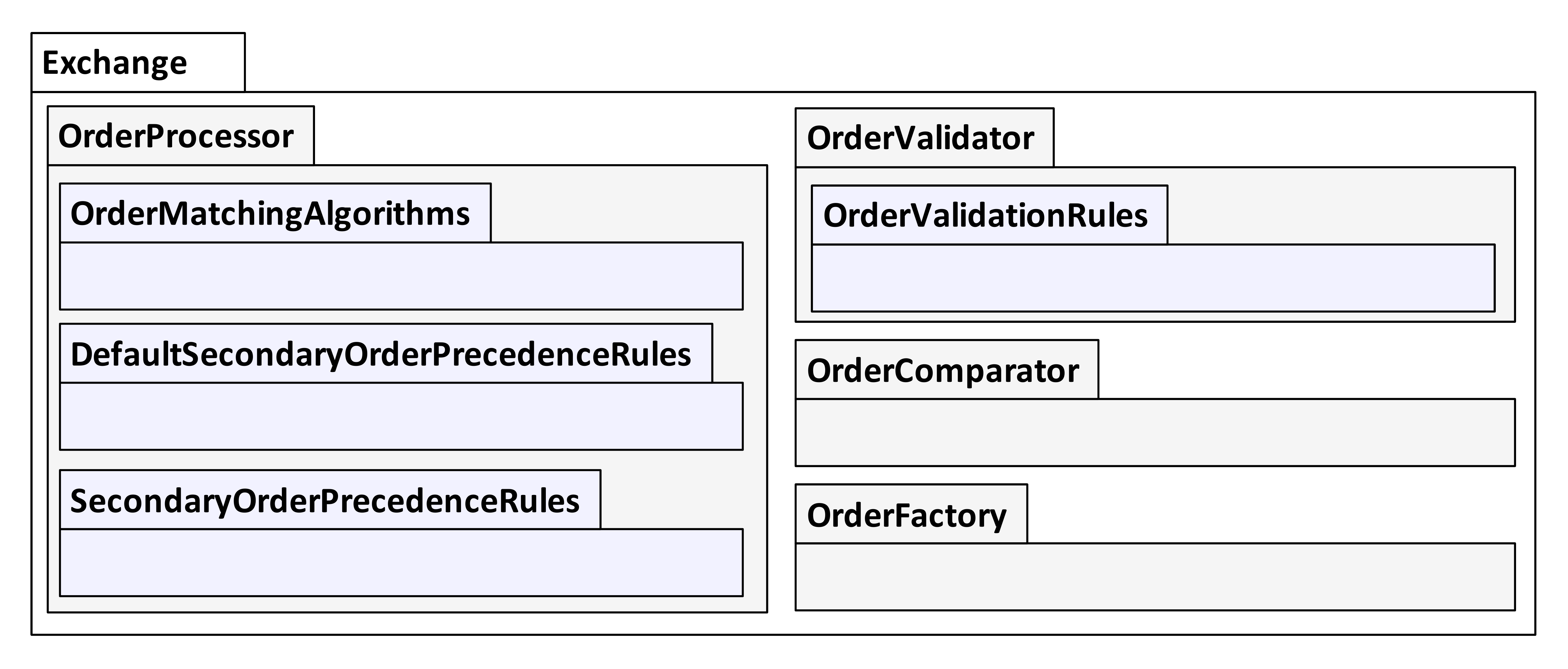}
  \caption{Exchange's software architecture}
  \Description{}
  \label{fig:exchangeArc}
\end{figure}

\begin{figure}[h]
  \centering
  \includegraphics[width=\columnwidth]{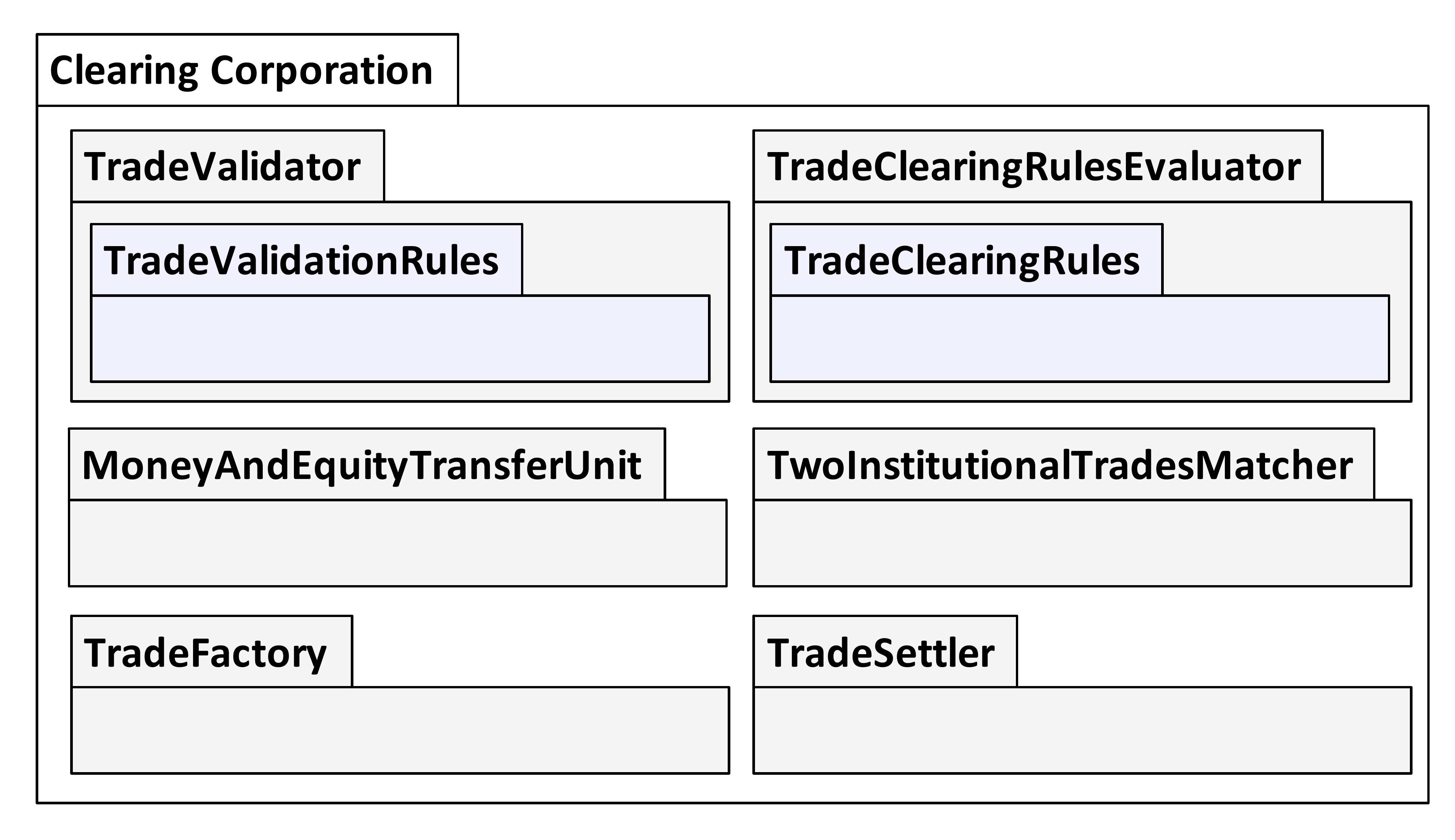}
  \caption{Clearing corporation's software architecture}
  \Description{}
  \label{fig:clearingCorpArc}
\end{figure}

This section provides the reader with information regarding different aspects of the architecture of our equity market SECO. We used Kruchten's software architecture development view to show the hierarchical decomposition of the system~\cite{kruchten1995architectural}.  Figure~\ref{fig:emSECOArc} shows the main view of the system that includes packages associated with each market participant. Figures ~\ref{fig:brokerArc}, ~\ref{fig:custodianArc}, ~\ref{fig:exchangeArc}, and ~\ref{fig:clearingCorpArc} exhibit the details of each package, namely broker, custodian, exchange, and clearing corporation. Packages colored in light purple represent variants.

Note that this implemented SECO is a small-scale and simplified version of a real operational one. Provided that an organization decides to develop a real SECO, it can make the most of this architecture's general ideas and make architectural decisions that work in a real environment. We have done the following activities to simplify our architecture and help the reader grasp ideas more easily.

\subsubsection{Simulating a distributed environment}
As building and simulating a SECO in a real distributed manner can add unnecessary complexities, we decided to place all market participants' software in a single code base. However, to make the software look distributed and enable it to be converted to a real distributed SECO with the least effort, we embedded a \textit{service registry} mechanism in all market participants' software. A service registry is a mechanism for establishing loosely-coupled connections between different units (or services) in a distributed environment. In our service registry, we used the market participants' object references instead of real web or IP addresses.

\subsubsection{Simplification of design by eliminating the reactive and recurring behavior}
In many real-world complex SECOs, a reactive mechanism is employed to make the system's behavior event-driven. Moreover, there are many recurring jobs in such systems for doing certain tasks required to be performed at specific intervals. Such behavior usually makes understanding and tracing of a software system difficult. Therefore, we decided to convert all reactive and recurring behavior to sequential one for simplicity. In our implementation, the reactive approach has been completely replaced with the sequential one so that all codes are executed sequentially. In addition, recurring methods have been tagged with the \textit{rec} suffix in the source code. Every time a recurring method should be triggered, we invoke them manually.

\subsection{An architectural Limitation}
As you can see in figure~\ref{fig:participantsRelations}, we have an option to have more than one instance of some market participants' software, such as custodians, brokers, exchanges, clearing banks, in our SECO. In this case, there can be multiple derived products of a certain participant's software in theory. In our proposed architecture, we avoided unnecessary complexities, coming from implementing a solution that creates and connects different products of a specific participant's software to the SECO. 

Although there are different instances of a certain market participant's software in the source code, for example, two brokers in a trading scenario, these instances are the same. This problem comes from the fact that products have been created from a set of similar class files, combined after compiling the source code. Thus, all of these class files lead to the generation of a single product, not multiple products. However, the mentioned limitation cannot be considered as a problem in a real SECO. When customers demand different products of a certain market participant's software, they receive a unique product and deploy it on their machine. Therefore, this problem does not happen in the real world.

\section{Application Realization}
In this section, we want to discuss our product derivation method and also some points on SECO implementation.

\subsection{Derivation of products}
We derive two different SECOs from our MPL, which are different in some features. To manage this stage properly, we benefited from the \textit{configuration map} feature of the FeatureIDE software plugin. Configuration map allows us to create a configuration file for each product and define our desired features in it. Two configuration files for our SECOs can be identified with the names \textit{Equity Market SECO A} and \textit{Equity Market SECO B} in the source code. 

\subsection{Variation points and variants' implementation approach}
The way variation points and variants are defined in software plays a principal role in the ease of the product derivation stage. To honor open-closed principle \cite{meyer1997object} \cite{martin2006agile}, for each market participants' software, we tried our best to implement each variant separately. To achieve this goal, instead of annotating different parts of a lengthy code section with some tags, we defined interfaces for variants to not only show variation points but also to make variation points and variants separate and distinguishable. We used FeatureIDE's \textit{Antenna} preprocessor \cite{antennaPreprocessor} to tag variants and making compile-time derivation possible.

\section{Validation of design}
To validate our work, we decided to test two derived SECOs from the previous section with three common order life cycle scenarios. Before doing so, it is a good idea to familiarize ourselves with these scenarios. In each scenario, all market participants' software pieces cooperate to support the order life cycle from the initial stage, order initiation, to the last stage, settlement.

\subsection{Institutional and retail orders' life cycles}
Institutional and retail order life cycles are two major order life cycles in the equity market. While the retail order life cycle can be easily understood by inspecting its respective diagram, the institutional one demands more explanation. 

\begin{figure}[h]
  \centering
  \includegraphics[width=\columnwidth]{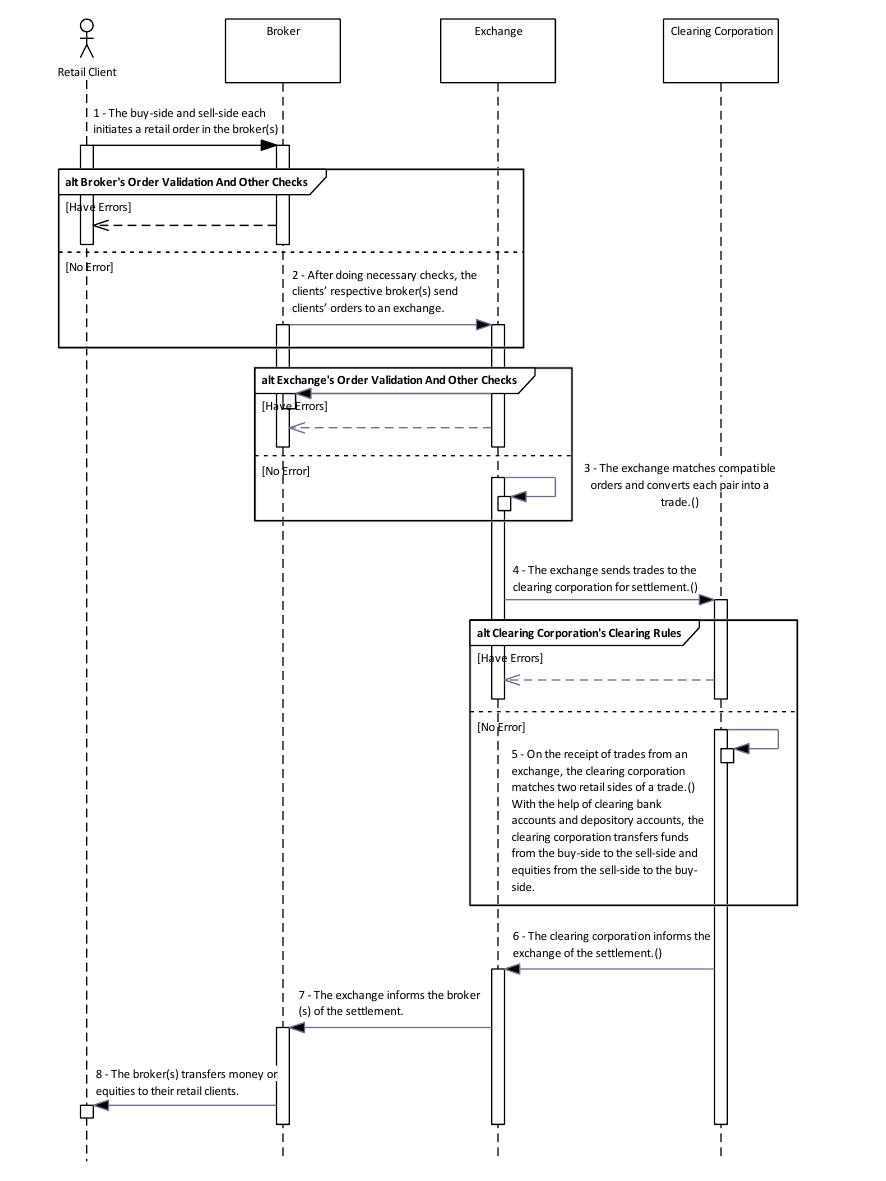}
  \caption{Retail order life cycle}
  \Description{}
  \label{fig:retailLifecycle}
\end{figure}

\begin{figure}[h]
  \centering
  \includegraphics[width=\columnwidth]{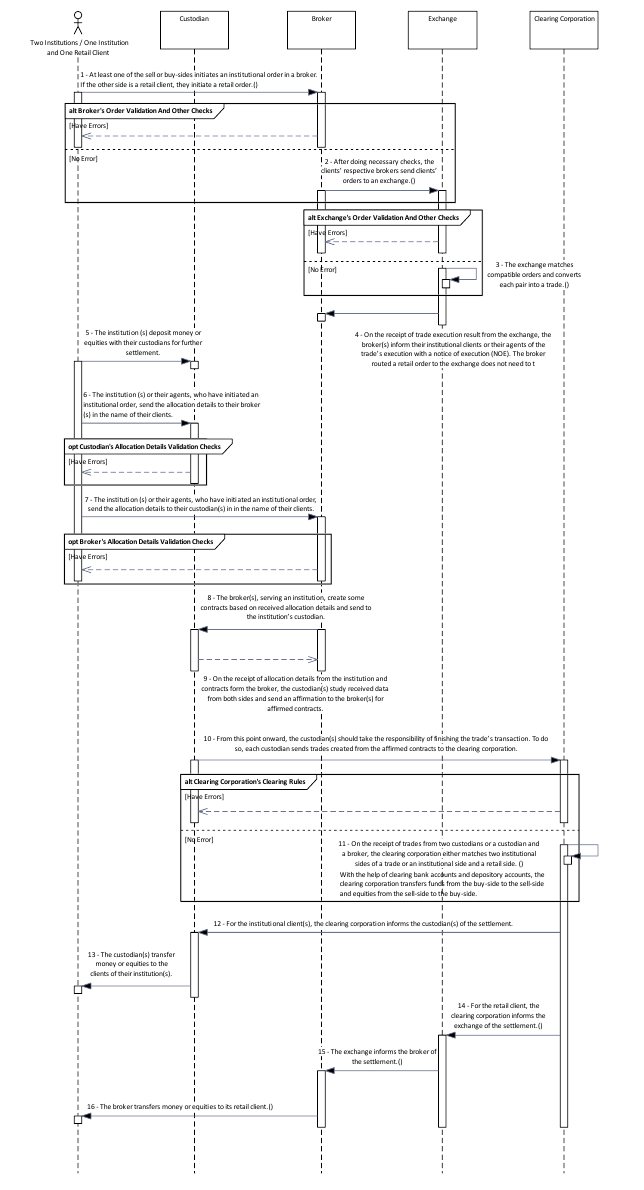}
  \caption{Institutional order life cycle}
  \Description{}
  \label{fig:institutionalLifecycle}
\end{figure}

In an institutional order life cycle scenario, a fund manager, who is the representative of an institution, wants to build a position immediately not to lose a good opportunity in the market for purchasing or selling equities. In this situation, the manager makes a large purchase or sell order in the broker. After the execution of the order by the broker, the manager allocates this large order to the institution's clients by creating a document in the name of each client, called \textit{allocation detail}. The manager then sends these allocation details to both the broker and custodian with that the institution works. The broker creates some \textit{contracts} from the provided allocation details and sends them to the institution's custodian. Since the custodian's responsibility is to protect its institutional client's interests, on receiving contracts from the broker, it checks them against its allocation details with some rules in a process called \textit{affirmation}. If these documents agree, the custodian sends affirmation to the broker and discharge the responsibility of settling trades from this point onward. For brevity, the rest of the process can be followed by the respective diagram. The life cycles of retail and institutional orders have been depicted in figure~\ref{fig:retailLifecycle} and figure~\ref{fig:institutionalLifecycle} and tabulated in table~\ref{tab:orderLifeCycleScenarios}. To best understand these scenarios, we made an example for each, which can be seen from the footnote \footnote{\href{https://github.com/KhosroPakmanesh/EquityMarketSoftwareEcosystem/blob/main/different order life-cycles with examples/Different order life-cycles with examples.pdf} Different order life cycles with examples}.

\begin{table*}
  \centering
  \caption{Order life-cyle scenarios}
  \label{tab:orderLifeCycleScenarios}
  \begin{tabular}{|p{3cm}|p{3cm}|p{11cm}|}
    \hline
    Client types&Life cycle type&Market participants\\
    \hline
    Two retail clients&Retail life cycle&Broker(s), exchange, clearing corporation, clearing bank(s), and depository\\
    \hline  
    A retail and an institutional client&Institutional life cycle&Broker(s), custodian(s), exchange, clearing corporation, clearing bank(s), and depository\\
    \hline  
    Two institutional clients&Institutional life cycle&Broker(s), custodian(s), exchange, clearing corporation, clearing bank(s), and depository\\
    \hline  
\end{tabular}
\end{table*}

\subsection{Implementing test scenarios}
To implement the scenarios, we used the data of the mentioned examples and designed three system tests to inspect the equity market SECO in action. As it was said, the whole process in each scenario is entirely automated, and there is no manual user intervention in the process except the times a user interacts with the SECO via a user interface, tagged with the \textit{ui} suffix in the source code, to provide the system with some necessary data to start the process, for example, making an order. 

In our system tests, we have described these scenarios step by step in detail, where the money and equities balance of every client is recorded as the process goes on. At the end of the system tests, by assessing each client's final money and equities balance, we have demonstrated that our equity market SECO works as expected. 

\section{Discussion of Results}
As we saw throughout this paper, the equity market, and generally, financial markets' SECOs can benefit profoundly from the MPL approach and unlock its potential. These SECOs are naturally rule-based software systems mainly composed of countless market rules, client instructions, government policies, and algorithms translated into different classifiable features. Since many of these features can be reused across different SECOs tailored to different countries' needs, the reusability that the MPL approach provides for the software development companies can be highly significant that mitigates the technical risks and enhances the chance of projects' success.

Since our work takes the readers to different stages and finally provides them with a practical and tangible source code, we believe that it dispels doubts regarding the vague aspects of applying this approach and contributes to minimizing technical risk to a large extent. Consequently, this research encourages them to consider this approach a workable and effective solution for their future projects. 

Further, one of the barriers to adopting an academic concept in the industry is the fact that converting theoretical ideas into practical solutions is a long and laborious process. An activity that can significantly reduce this barrier is providing industrial practitioners with a source code to help them understand academic concepts readily. Our work does such activity by presenting a source code that not only gives some examples and guidelines regarding the implementation of academic concepts but also offers the whole SECO that can be developed into a real one with some extra effort.

\section{Conclusion and Future Work}
In this paper, we explored the application of the MPL approach in the equity market SECO. We fully explained how SPLE approach guided the development of the necessary technical infrastructure for achieving STP. The domain engineering process of the equity market SECO began with extracting domain knowledge and modeling it. Then, we provided the reader with essential information about the equity market's participants and variability in market participants' software. Next, we discussed the architecture of the MPL. Following that, we entered the application engineering process, derived two SECOs from the MPL, and discussed some implementation details. Finally, we validated our approach by testing two SECOs and putting each under three system tests. This research revealed that the application of the MPL approach in the equity market SECO could yield promising results and deserve further studies.

Regarding future work, this research can be extended by four different activities. These activities are as follows:
\begin{itemize}
\item {Khanna describes that equity is the least complex asset to trade among other assets, such as derivatives. She believes that there is a direct correlation between an asset's complexity and the amount of possible automation for the trading of that asset \cite[pp.~40-42]{Khanna2008straight}. To put it simply, more complex assets can be traded with less automation. Therefore, one can work on more complex types of assets and design an MPL to automate the trading process of these assets.}
\item {One can consult with financial market experts from different countries to gain more information about various aspects of their markets. This activity helps the current SECO reflect and cover more similarities and differences and move towards becoming a mature MPL that supports the derivation of products that meet those countries' financial market needs.}
\item {As we mentioned previously, we left out some rules about fees a market participant should pay to another. Consequently, there is a tremendous opportunity to incorporate these and other rules into different market participants' software to make the equity market SECO more comprehensive and help practitioners gain even more realistic experience while simulating an order life cycle.}
\item{In the equity market's architecture section, we introduced a limitation about our work and then explained why in the deployment environment the purchasers of our SECO do not experience any problem. Still, in the simulation environment, we have the mentioned limitation. There is still room for suggesting a solution that removes this limitation and empowers the practitioners to connect different derived products to the SECO.}
\end{itemize}

\begin{acks}
We thank Yogesh Shetty and Samir Jayaswal for the order matching engine source code of their book \cite{shetty2006practicalsourcecode} that made this research possible.
\end{acks}

\bibliographystyle{ACM-Reference-Format}
\bibliography{manuscript}

\end{document}